\documentclass[aps,prb,twocolumn,superscriptaddress,showpacs]{revtex4}
\usepackage{eurosym}
\usepackage{amsfonts}
\usepackage{amssymb}
\usepackage{amsmath}
\usepackage{graphicx}
\usepackage{color}
\usepackage[hypertex,dvipdfm,colorlinks=true,urlcolor=blue,linkcolor=blue,citecolor=blue]{hyperref}% add hypertext capabilities
\begin{document}
\title{Manipulating superconductivity of $1T$-TiTe$_2$ by high pressure}
%Enhanced superconductivity of $1T$-TiTe$_2$ under uniaxial pressure: A first-principles prediction

\author{R. C. Xiao}
\affiliation{Key Laboratory of Materials Physics, Institute of Solid
State Physics, Chinese Academy of Sciences, Hefei 230031, China}
\affiliation{University of Science and Technology of China, Hefei, 230026, China}

\author{W. J. Lu}
\email{wjlu@issp.ac.cn}
\affiliation{Key Laboratory of Materials
Physics, Institute of Solid State Physics, Chinese Academy of
Sciences, Hefei 230031, China}

\author{D. F. Shao}
\affiliation{Key Laboratory of Materials Physics, Institute of Solid
State Physics, Chinese Academy of Sciences, Hefei 230031, China}

\author{J. Y. Li}
\affiliation{Key Laboratory of Materials Physics, Institute of Solid
State Physics, Chinese Academy of Sciences, Hefei 230031, China}
\affiliation{University of Science and Technology of China, Hefei, 230026, China}

\author{M. J. Wei}
\affiliation{Key Laboratory of Materials Physics, Institute of Solid
State Physics, Chinese Academy of Sciences, Hefei 230031, China}
\affiliation{University of Science and Technology of China, Hefei, 230026, China}

\author{H. Y. Lv}
\affiliation{Key Laboratory of Materials Physics, Institute of Solid
State Physics, Chinese Academy of Sciences, Hefei 230031, China}
\author{P. Tong}
\affiliation{Key Laboratory of Materials Physics, Institute of Solid
State Physics, Chinese Academy of Sciences, Hefei 230031, China}
\author{X. B. Zhu}
\affiliation{Key Laboratory of Materials Physics, Institute of Solid
State Physics, Chinese Academy of Sciences, Hefei 230031, China}
\author{Y. P. Sun}
\email{ypsun@issp.ac.cn}
\affiliation{High Magnetic Field Laboratory, Chinese Academy of
Sciences, Hefei 230031, China}
\affiliation{Key Laboratory of Materials Physics, Institute of Solid
State Physics, Chinese Academy of Sciences, Hefei 230031, China}
\affiliation{Collaborative Innovation Center of Microstructures,
Nanjing University, Nanjing 210093, China }

\begin{abstract}
Superconductivity of transition metal dichalcogenide $1T$-TiTe$_2$ under high pressure was investigated by the first-principles calculations. Our results show that the superconductivity of $1T$-TiTe$_2$ exhibits very different behavior under the hydrostatic and uniaxial pressure. The hydrostatic pressure is harmful to the superconductivity, while the uniaxial pressure is beneficial to the superconductivity. Superconducting transition temperature $T_C$ at ambient pressure is 0.73 K, and it reduces monotonously under the hydrostatic pressure to 0.32 K at 30 GPa. While the $T_C$ increases dramatically under the uniaxial pressure along $c$ axis. The established $T_C$ of 6.34 K under the uniaxial pressure of 17 GPa, below which the structural stability maintains, is above the liquid helium temperature of 4.2 K. The increase of density of states at Fermi level, the redshift of $F(\omega)$/$\alpha^2F(\omega)$ and the softening of the acoustic modes with pressure are considered as the main reasons that lead to the enhanced superconductivity under uniaxial pressure. In view of the previously predicted topological phase transitions of $1T$-TiTe$_2$ under the uniaxial pressure [Phys. Rev. B \textbf{88}, 155317 (2013)], we consider $1T$-TiTe$_2$ as a possible candidate in transition metal chalcogenides for exploring topological superconductivity.
\end{abstract}
 %, 74.25.Kc
\maketitle
%===========================================================%

\section{Introduction}
Transition metal dichalcogenides (TMDCs) MX$_2$, with M a transition metal (e.g. M = Ti, Mo, Ta, W) and X a chalcogen atom (S, Se, Te), is an emerging family of layered materials. The intra layer is composed by X-M-X sandwich structure, which is attracted by the van der Waals forces between the inter layers. TMDCs material that can be semiconductor, metal, charge density wave (CDW) system, or superconductor has become a rich playground to discover new materials with diverse physical phenomena and properties. Broad application prospects of TMDCs such as transistors, photodetectors, electroluminescent devices,\cite{Wang_Q_H_2012} van der Waals heterostructures with high on/off current ratio,\cite{Geim_A_K_2013} and topological field-effect transistors based on quantum spin Hall effect,\cite{Qian_Xiaofeng_2014} have been triggered great attention.

MTe$_2$ is a typical material with rich physical properties and exhibits unique properties compared with MS$_2$/MSe$_2$ due to the strong $p$-$d$ hybrid and spin-orbit coupling effect. For instance, the competition between the charge/orbital density wave (CDW/ODW) and superconductivity was observed in IrTe$_2$;\cite{Yang_J_J_2012,Kamitani_M_2013} topological Dirac point was found in the HfTe$_2$/AlN epitaxial system;\cite{Aminalragia-Giamini_2016} large and non-saturating magnetoresistance,\cite{Ali_M_N2014,Pletikos_2014,Keum_Dong_Hoon_2015} type-II Weyl points,\cite{Soluyanov_2015,Yan_Sun 2015,Deng_Ke_2016,Wang_Z_2016} and pressure driven superconductivity\cite{Pan_X_C_2015,Kang_D_2015,Qi_Yanpeng_2016} were observed in WTe$_2$ and MoTe$_2$; topologically nontrivial surface state with Dirac cone was found in PdTe$_2$ superconductor;\cite{Liu_Yan_2015} and the type-II Dirac Fermions in PtTe$_2$\cite{Huang_Huaqing_2016,Mingzhe_Yan_2016} was recently theoretically proposed and confirmed by experiment.

$1T$-TiTe$_2$ is a textbook Fermi-liquid system.\cite{Claessen_1992,Claessen_1996} Though it is a simple physical system, abundant physical phenomena can be realized via various kinds of manipulation. Anomalous electron transport was found in the back-gated field-effect transistors with $1T$-TiTe$_2$ thin-film channels.\cite{Khan_J_2012} Large negative magnetoresistance was reported in two-dimensional spin-frustrated $1T$-TiTe$_{2-x}$I$_x$.\cite{Guo_Y_2014} Bulk and monolayer $1T$-TiS$_{2-x}$Te$_x$ show topological phases under certain concentration of S/Te.\cite{Zhu_Zhiyong_2013} $1T$-TiTe$_2$ was recently predicted to undergo series of topological phase transitions under high pressure.\cite{Zhang_Qingyun_2013} $1T$-TiTe$_2$ is a semimetal with an overlap of valence and conduction bands of 0.6 eV,\cite{de_Boer_1984} however the superconductivity has not been experimentally found down to 1.1 K at ambient pressure.\cite{Allen_1994}

Pressure, an important controllable parameter that can effectively tune the lattice structures and the corresponding band structure, has become an effective way to introduce superconductivity and study the relationship between the superconductivity and other physical phenomena in TMDCs. For instance, the metallization and the highest onset $T_C$ of 11.5 K were realized in MoS$_2$ under high pressure;\cite{Chi_Z_H_2014,Chi_Zhenhua_2015} the relationship between CDW and superconductivity in $1T$-TaS$_2$\cite{Ritschel_T_2013,Sipos_B_2008} and $1T$-TiSe$_2$\cite{Kusmartseva_A_F_2009,Calandra_2011} was studied under high pressure; the pressure driven superconductivity and suppressed magnetoresistance were observed in WTe$_2$\cite{Pan_X_C_2015,Kang_D_2015} and MoTe$_2$.\cite{Qi_Yanpeng_2016,ChenF_C_2016}

In this work, we focus on the possibility of superconductivity of $1T$-TiTe$_2$ under high pressure by first-principles calculations. Our results show that the $T_C$ at ambient pressure is 0.73 K, and it reduces under the hydrostatic pressure to 0.32 K at 30 GPa. While the uniaxial pressure along $c$ axis can increase the $T_C$ dramatically to 6.34 K maximally. We explained the different behavior of superconductivity under the hydrostatic/uniaxial pressure based on the varieties of electronic/phonon structure and the electron-phonon coupling effect.
\section{Methods}
The first-principles calculations based on density functional theory (DFT) were carried out using QUANTUM-ESPRESSO package.\cite{Paolo_Giannozzi_2009} The ultrasoft pseudo-potentials and the local density approximation (LDA) according to the PZ functional were used. The energy cutoff for the plane wave (charge density) basis was set to 35 Ry (350 Ry). The Brillouin zone (BZ) was sampled with a $16\times16\times8$ mesh of \emph{\textbf{k}}-points. The Vanderbilt-Marzari Fermi smearing method with a smearing parameter of $\sigma = 0.02$ Ry was used. The lattice constants and ions were optimized using Broyden-Fletcher-Goldfarb-Shanno (BFGS) quasi-newton algorithm. Electronic properties are calculated including the spin-orbit coupling effect. Since the spin-orbit coupling effect is less important in describing the vibrational properties,\cite{Chis-prb-2012,Verstraete-prb-2008} the calculation of phonon dispersion is carried out neglecting this effect. The phonon dispersion and electron-phonon coupling constants were calculated using density functional perturbation theory (DFPT)\cite{Baroni_Stefano_2001} with an $8\times8\times4$ mesh of \emph{\textbf{q}}-points. The double Fermi-surface averages of electron-phonon matrix elements were calculated using a $32\times32\times16$ mesh of \emph{\textbf{k}}-points.
\section{Results and discussion}
$1T$-TiTe$_2$ has a layered structure with space group $P$\=3m1 ($1T$-CdI$_2$ structure), with one Ti atom and two Te atoms located at (0, 0, 0) and (1/3, 2/3, $\pm z$) sites, respectively. The crystal structure and BZ are displayed in Fig. 1. The optimized lattice parameters are 3.677 $\mathrm\AA$ and 6.331 $\mathrm\AA$ for $a$ and $c$ respectively. The lattice parameters are slightly underestimated by $2.4 \%$ compared with the experimentally obtained ones,\cite{Claessen_1996,Arnaud_1981} and such underestimation exists normally in the LDA calculations.
As expected, when the hydrostatic pressure is applied the lattice is suppressed, as shown in Fig. 2(a). Due to the weak van der Waals coupling between adjacent layers, the reduction of $c$ is more substantial than that of $a$. By fitting the pressure-energy data to the Birch-Murnaghan equation of state, the bulk modulus $B_0$ and its derivative $B_0'$ for $1T$-TiTe$_2$ were calculated to be 51.1 GPa and 4.54 respectively.
\begin{figure}[h]
\begin{flushleft}
\includegraphics[width=0.9\columnwidth]{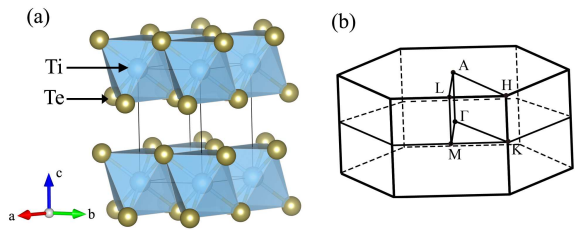}
\caption{(Color online) (a) Crystal structure and (b) Brillouin zone of $1T$-TiTe$_2$.}
\end{flushleft}
\end{figure}

\begin{figure}[h]
\begin{flushleft}
\includegraphics[width=0.9\columnwidth]{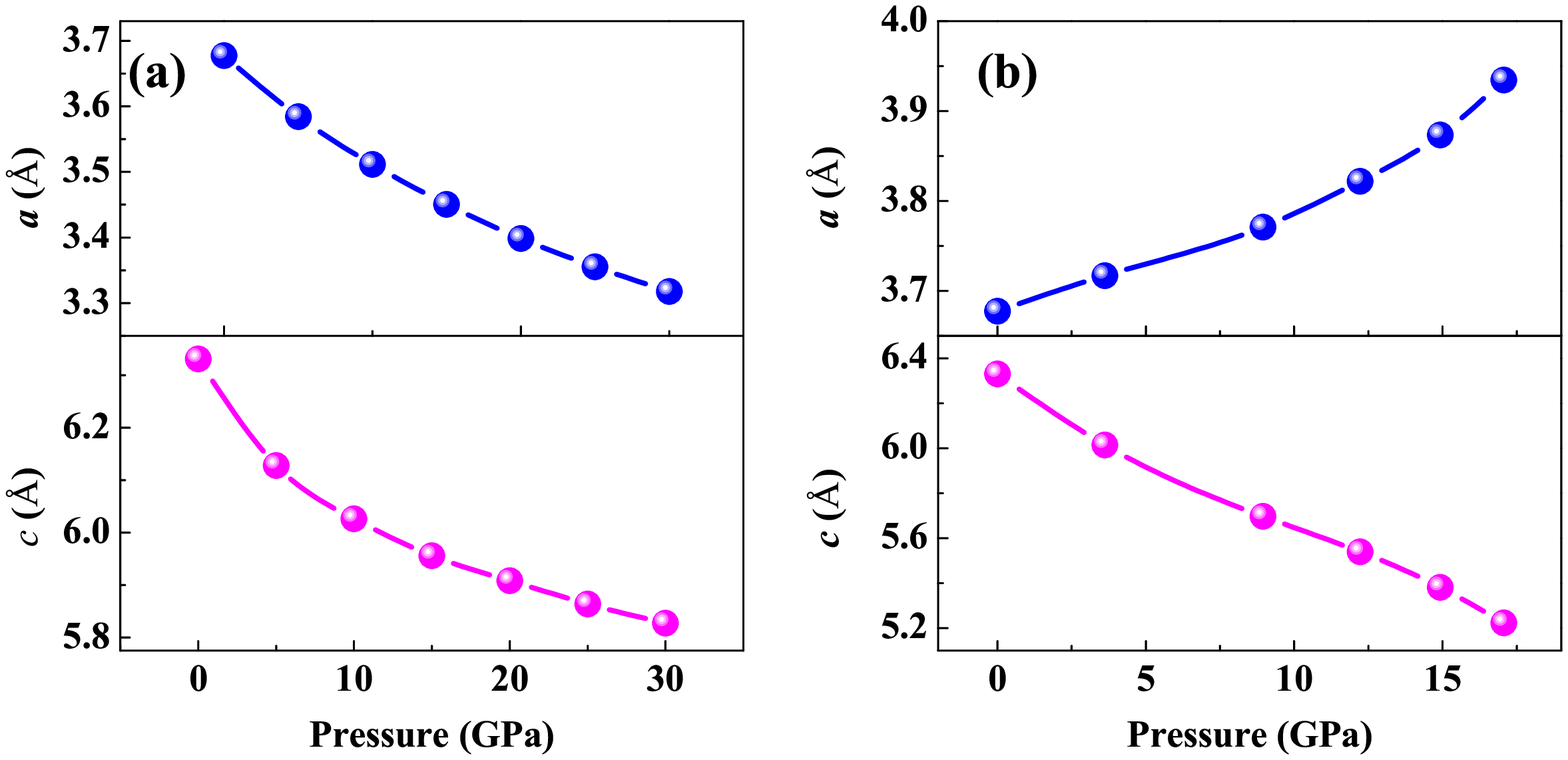}
\caption{(Color online) Lattice parameters under the (a) hydrostatic pressure and (b) uniaxial pressure along $c$ axis.}
\end{flushleft}
\end{figure}
The calculated electronic structure of $1T$-TiTe$_2$ at ambient pressure is shown in Fig. 3(c), which is reasonable agreement with the previously experimental and theoretical reports.\cite{de_Boer_1984,Reshak_Ali_Hussain_2003} The density of states (DOS) near the Fermi level ($E_F$) are mostly contributed by the Ti $d$ and Te $p$ orbitals. The electronic structure shows the semimetal feature with the DOS at Fermi level ($N(E_F)$) of 1.6 states/eV. The valence band maximum (VBM) and conduction band minimum (CBM) are located at the $\Gamma$ and L points, respectively.
The calculated phonon dispersion is shown in Fig. 3(d). The irreducible representations of the $\Gamma$ point phonons are $\Gamma= E_g + A_{1g} + 2E_u + 2A_{2u}$, and the corresponding optical vibration modes are illustrated in Fig. 4(a). The $E_g$ and $A_{1g}$ are Raman active modes, and the calculated frequencies of 105/150 $cm^{-1}$ for $E_g$/$A_{1g}$ are very close to 102/145 $cm^{-1}$ observed in experiment.\cite{Hangyo_Masanori_1983}

\begin{figure*}
\includegraphics[width=0.9\textwidth]{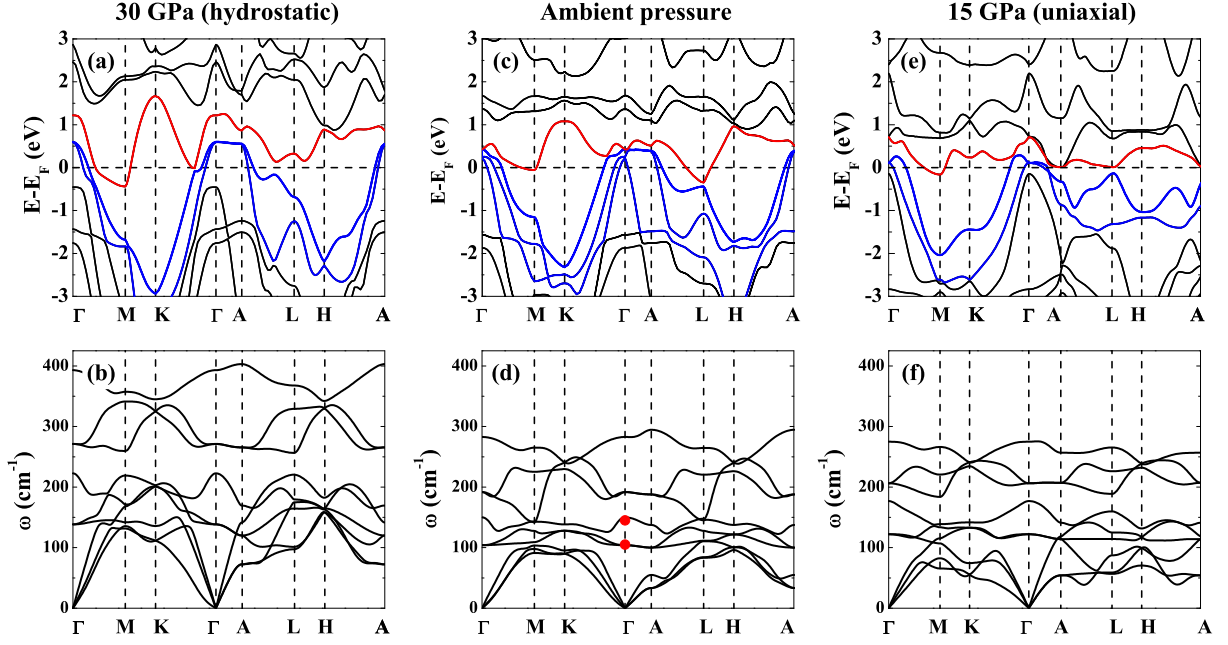}
\caption{(Color online) Electronic structures and phonon dispersions of $1T$-TiTe$_2$ under the (a, b) hydrostatic pressure of 30 GPa, (c, d) ambient pressure and (e, f) uniaxial pressure of 15 GPa. The red dots in (d) denote the Raman frequencies observed in experiment (Ref. 41). For the convenience of comparison, the valence and conduction bands crossing the $E_F$ decorated by blue and red respectively.}
\end{figure*}

\begin{figure}[h]
\includegraphics[width=0.95\columnwidth]{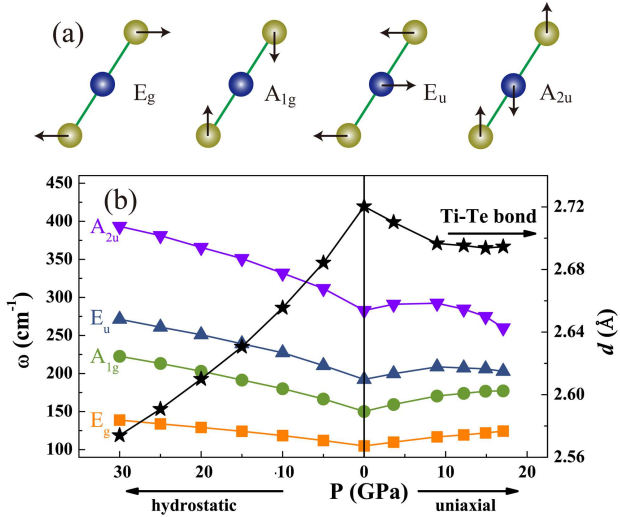}
\caption{(Color online) (a)Schematic illustration of the optical vibration modes at $\Gamma$ point, Ti and Te are denoted by blue and dark yellow balls, respectively. (b)Phonon frequencies at $\Gamma$ point and Ti-Te bond length under the hydrostatic and uniaxial pressure.}
\end{figure}

Applying pressure reduces the lattice parameters and enhances the atom interaction, making the band structure more dispersive and increasing the overlap of the valence bands and conduction bands. The electronic structure and phonon dispersion under the hydrostatic pressure of 30 GPa are shown in Figs. 3(a) and (b), where the CBM of electronic structure is changed from the L point to the M point.
With the lattice parameters and Ti-Te bond length decreasing, the frequencies of $\Gamma$ phonons increase monotonously under the pressure as demonstrated in Fig. 4(b), and especially the variation of $A_{2u}$ is more evidently. Meanwhile, the phonons of the whole BZ also shift to higher frequency with the hydrostatic pressure, as illustrated for the phonon density of states $F(\omega)$ in Fig. 6(a).

Though the band structures are more dispersive, the $N(E_F)$ decreases with the hydrostatic pressure (Fig. 5(a)). To reveal the pressure effect on the orbitals at $E_F$, the partial DOS at $E_F$ ($N^{P}(E_F)$) is shown in Figs. 5(b) and (c). The $N^{P}(E_F)$ of Ti $d$ decreases dramatically under the uniaxial pressure (Fig. 5(b)), and the $N^{P}(E_F)$ of Te $p_x$+$p_y$ and $p_z$ increase and decrease respectively under the hydrostatic pressure (Fig. 5(c)). The charge density at $E_F$ on (110) plane under the hydrostatic pressure of 30 GPa and ambient pressure are also shown in Figs. 5(g) and (h).

We estimated the superconducting transition temperature $T_C$ based on the Allen-Dynes-modified McMillan equation\cite{Allen_P_B1975}
\begin{equation}
T_c=\frac{\omega_{log}}{1.2}\exp\left(-\frac{1.04(1+\lambda)}{\lambda-\mu^*-0.62\lambda\mu^*}\right) ,
\end{equation}
where the Coulomb pseudopotential $\mu^*$ is set to a typical value of 0.1.\cite{Calandra_2011,Penev_E_S_2016,Kortus_J_2001} The logarithmically averaged characteristic phonon frequency $\omega_{log}$ is defined as
\begin{equation}
\omega_{log}=\exp\left(\frac{2}{\lambda}\int\frac{d\omega}{\omega}\alpha^{2}F(\omega)\log\omega\right).
\end{equation}
The total electron-phonon coupling constant $\lambda$ can be obtained by
\begin{equation}
\lambda=\sum_{\mathbf{q} v}\lambda_{\mathbf{q} v}
=2\int\frac{\alpha^2F(\omega)}{\omega}\mathrm{d}\omega ,
\end{equation}
where the Eliashberg spectral function is
\begin{equation}
  \alpha^2F(\omega)=\frac{1}{2\pi N(E_F)}\sum_{\mathbf{q}v}
  \delta(\omega-\omega_{\mathbf{q}v})\frac{\gamma_{\mathbf{q}v}}{\hbar \omega_{\mathbf{q}v}}.
\end{equation}

The $T_C$ at ambient pressure is calculated to be 0.73 K, coinciding with the fact that the superconductivity was not found above 1.1 K in the experiment.\cite{Allen_1994} As discussed above, applying hydrostatic pressure increases the phonon frequencies, therefore the Debye temperature raises, so does the $\omega_{log}$ (see Fig. 7(a)). Similar to the $F(\omega)$, Eliashberg function $\alpha^2F(\omega)$ shifts to higher frequency with the increase of hydrostatic pressure (Fig. 6(b)). Therefore, according to Eq. (3), the $\lambda$ decreases with the hydrostatic pressure (see Fig. 7(b)). As a result, the $T_C$ decreases monotonously with the pressure from 0.73 K at ambient pressure to 0.32 K at 30 GPa (Fig. 7(c)), indicating that the hydrostatic pressure does not benefit to introduce the experimentally detected superconductivity, which is unlike the cases of emerging superconductivity in the semimetal WTe$_2$ and MoTe$_2$\cite{Pan_X_C_2015,Kang_D_2015,Qi_Yanpeng_2016,ChenF_C_2016} under hydrostatic pressure.

\begin{figure}%[h]
\includegraphics[width=1\columnwidth]{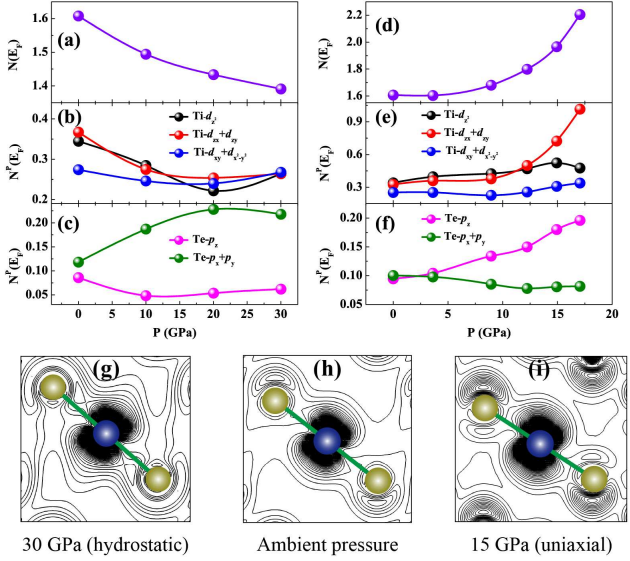}
\caption{(Color online) $N(E_F)$ and $N^{P}(E_F)$ (in states/eV) under the (a)-(c) hydrostatic and (d)-(f) uniaxial pressure. The charge density at $E_F$ on (110) plane under the (g) hydrostatic pressure of 30 GPa, (h) ambient pressure and (i) uniaxial pressure of 15 GPa. The difference between two contour lines in (g)-(i) is set to the same. Ti and Te are denoted by blue and dark yellow balls, respectively.}
\end{figure}

Due to the layer structure, uniaxial pressure along $c$ axis can be easily applied in experiment. Therefore the superconductivity under the uniaxial pressure along $c$ axis is studied as well in our research. When the uniaxial pressure along $c$ axis is applied, the $c$ axis reduces meanwhile $a$ axis expands (Fig. 2(b)). The variations of $a$ and $c$ are more evidently than that under the hydrostatic pressure. The average Poisson's ratio $v=-\Delta a/\Delta c$ in the range of our study is 0.23. The electronic structure is less dispersive and the overlap is reduced with the uniaxial pressure. The electronic structure under the uniaxial pressure of 15 GPa is shown in Fig. 3(e), where the CBM is changed from the L point to the M point. The phonon dispersion under the uniaxial pressure of 15 GPa is shown in Fig. 3(f), where the acoustic modes soften. %where there are dips in the acoustic modes, indicating the softening of phonon.

As demonstrated in Fig. 4(b), with the increase of uniaxial pressure, both $E_g$ and $A_{1g}$ modes that only involve the vibrations of Te atom increase monotonously. While, the $E_u$ and A$_{2u}$ modes increase at first and then decrease as the pressure larger than 8.9 GPa, which could be attributed to the combination of the expansion of $ab$ plane and the slow decrease of the Ti-Te length (see Fig. 4(b)). Meanwhile, the variation of $E_u$ and A$_{2u}$ under the uniaxial pressure is less significant than that under the hydrostatic pressure due to the fact that the reduction of Ti-Te length is much less than that under the hydrostatic pressure.

\begin{figure}%[h]
\includegraphics[width=1\columnwidth]{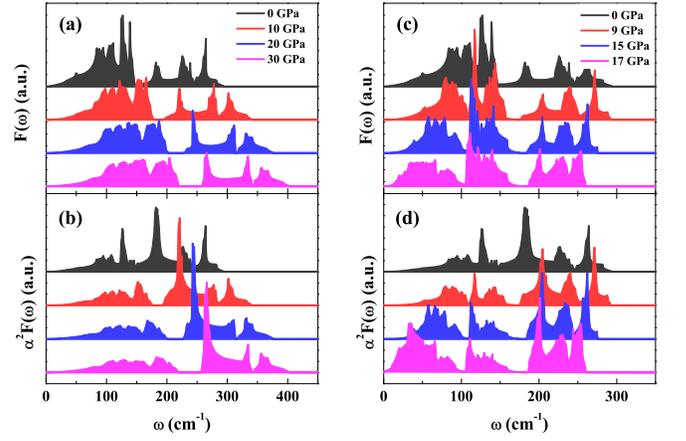}
\caption{(Color online) (a)/(c) Phonon density of states $F(\omega)$ and (b)/(d) Eliashberg function $\alpha^2F(\omega)$ under hydrostatic pressure (left) and uniaxial pressure (right).}
\end{figure}

\begin{figure}%[h]
\includegraphics[width=0.9\columnwidth]{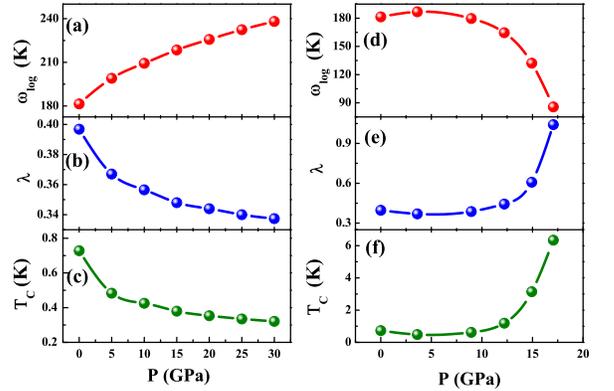}
\caption{(Color online) Calculated $\lambda$, $\omega_{log}$ and $T_C$ of $1T$-TiTe$_2$ under the (a)-(c) hydrostatic and (d)-(f) uniaxial pressure.}
\end{figure}
The uniaxial pressure raises the $N(E_F)$ very much (see Fig. 5(d)). The $N^{P}(E_F)$ of Ti $d_{zx}$+$d_{zy}$ increases dramatically under the uniaxial pressure (Fig. 5(e)), and the trend of $N^P(E_F)$ of Te $p_x$+$p_y$ and $p_z$ under the uniaxial pressure (Fig. 5(f)) is opposite to that under the hydrostatic pressure. The increase of the $N^{P}(E_F)$ of Te $p_z$ and Ti $d_{zx}$+$d_{zy}$ will increase the orbital overlap of Ti-Te atoms at $E_F$ as shown in Fig. 5(i), while the orbital overlap is not such case under the hydrostatic pressure (Fig. 5(g)). The strong $\sigma$ bands crossing the $E_F$ due to the orbital overlap at $E_F$ is one of the main reasons of high $T_C$ of superconductors MgB$_2$\cite{Kortus_J_2001,An_J_M_2001,Singh_2006} and H$_3$S.\cite{Drozdov_2015,Bernstein_2015} %We consider that the enhanced orbital overlap of Ti-Te atoms at $E_F$ under the uniaxial pressure could also contribute to the enhanced superconductivity.

As shown in Fig. 6(c), even though some optical phonon $F(\omega)$ peaks shift to higher frequency in some cases, the overall $F(\omega)$ shifts to lower frequency with the increase of uniaxial pressure, especially for the acoustic phonons and the $\omega_{log}$ decreases with uniaxial pressure as well (Fig. 7(d)). Similar to the $F(\omega)$, $\alpha^2F(\omega)$ shifts to lower frequency (Fig. 6(d)), and the proportion of its low frequency part increases with the uniaxial pressure duo to the softening of the acoustic modes. According to Eq. (3), the mode with lower frequency and the larger $\alpha^2F(\omega)$ will strongly contribute to the electron-phonon coupling. Therefore, as shown in Fig. 7(e), the $\lambda$ basically increases with the increase of uniaxial pressure. Considering overall effects, the $T_C$ changes slowly at first while it increases dramatically as the uniaxial pressure larger than 8.9 GPa (Fig. 7(f)). The $N(E_F)$ decreases with the hydrostatic pressure, and increases dramatically under the uniaxial pressure (Figs. 5(a) and (d)). The varying trend of $T_C$ with pressure coincides with that of $N(E_F)$ (Figs. 7 (c) and (f)). This result is consistent with the scenario that, as a general rule of BCS, larger $N(E_F)$ is in favor of higher $T_C$.

Superconductivity with a relatively high $T_C$ often emerges in the vicinity of structural instability. Just beneath the structural instability, the $T_C$ of 6.34 K under the uniaxial pressure of 17 GPa is estimated, above the liquid helium temperature of 4.2 K. Under higher uniaxial pressure, the structure is no longer stable, which is estimated from the calculated imagine frequency in the acoustic modes. Our results show that the hydrostatic pressure is harmful to the superconductivity, while the uniaxial pressure is beneficial to the superconductivity of $1T$-TiTe$_2$.

\begin{figure}%[h]
\includegraphics[width=1.0\columnwidth]{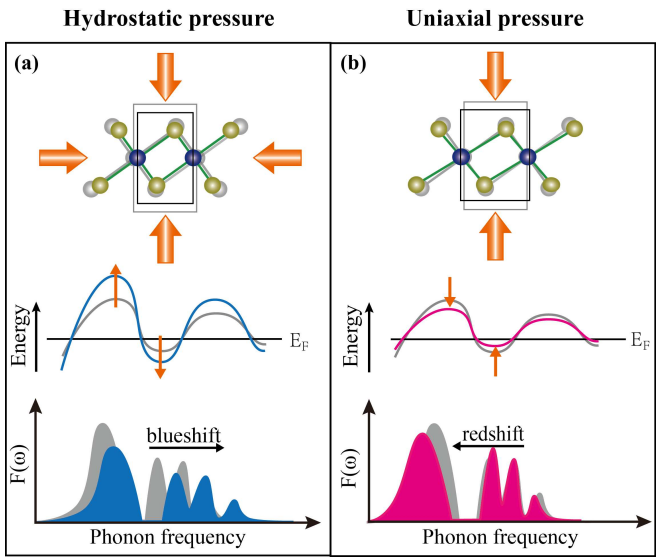}
\caption{(Color online) Schematic diagram of the effects of (a) hydrostatic pressure and (b) uniaxial pressure on the crystal (top), electronic (middle) and phonon (bottom) structures. The gray symbols in each graph denote the corresponding states at ambient pressure. Ti and Te are denoted by blue and dark yellow balls, respectively.}
 %and shifted to the middle of $c$ here.}
\end{figure}

We draw a schematic diagram to describe the effects of hydrostatic pressure and uniaxial pressure on the crystal, electronic and phonon structures (Fig. 8). Both $a$, $c$ and Ti-Te bond are compressed under the hydrostatic pressure (Fig. 8(a)). The stronger band dispersion makes the $N(E_F)$ reduce. The phonon density of states $F(\omega)$ shifts to higher frequency. The decrease of $N(E_F)$ and the blueshift of $F(\omega)$ are not in favor of superconductivity, as discussed above. While under the uniaxial pressure $c$ is suppressed and $a$ is expanded accordingly (Fig. 8(b)). The band dispersion becomes weaker, therefore the $N(E_F)$ increases with the increase of uniaxial pressure. The acoustic phonon modes soften, and $F(\omega)$ shifts to lower frequency. These two factors (the increase of $N(E_F)$ and the redshift of $F(\omega)$) are in favor of superconductivity, as discussed above. We think this physical scenario is not only applicable to $1T$-TiTe$_2$, but also to other layered semimetal TMDC materials. We propose that the uniaxial pressure can provide an alternative method for enhancing or finding superconductivity in TMDCs.

The previous investigation shows that $1T$-TiTe$_2$ is topological trivial at ambient pressure, but it was predicted to undergo series of topological phase transitions under pressure, which is related to the band inversions at different points of the BZ.\cite{Zhang_Qingyun_2013} Therefore applying the uniaxial pressure, one can expect to obtain the topological phase and the enhanced superconductivity in $1T$-TiTe$_2$ at the same time. As suggested, introducing superconductivity into the topological material could make them to be topological superconductor,\cite{Hasan_M_Z_2010,Qi_Xiao_Liang_2011} which has a full pairing gap in the bulk and a gapless surface state consisting of Majorana fermions. The possible topological superconductivity in $1T$-TiTe$_2$ under pressure is needed to be studied in the further experimental and theoretical studies.
\section{Conclusion}
Using the first-principles calculations, we demonstrated that the superconductivity of $1T$-TiTe$_2$ is suppressed under the hydrostatic pressure and enhanced under the uniaxial pressure. The increase of $N(E_F)$, the redshift of $F(\omega)$/$\alpha^2F(\omega)$ and the softening of the acoustic phonon modes with the uniaxial pressure contribute to the enhanced superconductivity. When the uniaxial pressure of 17 GPa is applied, the maximum $T_C$ of 6.34 K in our research is obtained. The uniaxial pressure provides an alternative method to manipulate superconductivity in TMDCs. Under reasonable pressure, the topological state and superconductivity may appear at the same time in $1T$-TiTe$_2$. The superconductivity and topological property in $1T$-TiTe$_2$ under pressure will expand its physics and applications.
%The uniaxial pressure provides an alternative method to manipulate superconductivity in TMDCs.
\begin{acknowledgments}
This work was supported by the National Key Research and Development Program of China under Contract No. 2016YFA0300404, the National Nature Science Foundation of China under Contract Nos. 11674326, 11404340, 11274311, 11404342 and U1232139, Youth Innovation Promotion Association of CAS (2012310), Key Research Program of Frontier Sciences of CAS (QYZDB-SSW-SLH015) and Hefei Science Center of CAS (2016HSC-IU011). The calculations were partially performed at the Center for Computational Science, CASHIPS.
\end{acknowledgments}

\end {document}